\documentclass[aps,prl,reprint,showpacs,amssymb,amsmath,superscriptaddress]{revtex4-1}

\usepackage{graphicx}
\usepackage{color}
\usepackage{times}
\usepackage{ulem}
\usepackage{verbatim}

\newcommand{\mean}[1]{\expec{#1}}
\newcommand{\Rb}{$^{87}$Rb }

\newcommand{\CS}{\mathrm{CS}}
\newcommand{\sigsum}{\Sigma_S}

\newcommand{\ket}[1]{\lvert#1\rangle}
\newcommand{\expec}[1]{\langle #1 \rangle}

\newcommand{\comm}[1]{}

\def\del#1{\comm{#1}}

\begin{document}
\title{Preparation of ultracold atom clouds at the shot noise level}

\author{M. Gajdacz}\altaffiliation{These authors contributed equally to this work. Corresponding author: hilliard@phys.au.dk}
\affiliation{Institut for Fysik og Astronomi, Aarhus Universitet, Ny Munkegade 120, 8000 Aarhus C, Denmark.}
\author{A. J. Hilliard}
\altaffiliation{These authors contributed equally to this work. Corresponding author: hilliard@phys.au.dk}
\affiliation{Institut for Fysik og Astronomi, Aarhus Universitet, Ny Munkegade 120, 8000 Aarhus C, Denmark.}
\author{M. A. Kristensen}
\affiliation{Institut for Fysik og Astronomi, Aarhus Universitet, Ny Munkegade 120, 8000 Aarhus C, Denmark.}
\author{P. L. Pedersen}
\affiliation{Institut for Fysik og Astronomi, Aarhus Universitet, Ny Munkegade 120, 8000 Aarhus C, Denmark.}
\author{C. Klempt}
\affiliation{Institut f\"ur Quantenoptik, Leibniz Universit\"at Hannover, Welfengarten 1, D-30167 Hannover, Germany.}
\author{J. J. Arlt}
\affiliation{Institut for Fysik og Astronomi, Aarhus Universitet, Ny Munkegade 120, 8000 Aarhus C, Denmark.}
\author{J. F. Sherson}
\affiliation{Institut for Fysik og Astronomi, Aarhus Universitet, Ny Munkegade 120, 8000 Aarhus C, Denmark.}


\begin{abstract}

We prepare number stabilized ultracold atom clouds  through the real-time analysis of non-destructive images and the application of feedback. In our experiments, the atom number ${N\sim10^6}$ is determined by high precision Faraday imaging with uncertainty $\Delta N$ below the shot noise level, i.e., $\Delta N <\sqrt{{N}}$. Based on this measurement, feedback is applied to reduce the atom number to a user-defined target, whereupon a second imaging series probes the number stabilized cloud. By this method, we show that  the atom number in ultracold clouds can be prepared below the shot noise level.

\end{abstract}
\maketitle

{
Over the past decade, experiments with ultracold atomic samples have matured from the proof-of-concept level to a development platform for  technologies such as 
quantum sensors and quantum simulators. One rapidly expanding technique 
is the manipulation of quantum systems using measurements and feedback~\cite{Zhang2014,Sayrin2011,Briegel2009}. To limit the back-action, usually a `weak' measurement is employed, such as detecting the phase shift induced by an atomic ensemble 
on an off-resonant laser beam~\cite{Hope2004}. Recent experiments 
have demonstrated feedback control of motion in an optical lattice~\cite{Morrow2002}, a quantum memory for light~\cite{Julsgaard2004},  deterministic spin squeezing~\cite{PhysRevLett.116.093602}, stabilization of an atomic system against decoherence~\cite{Vanderbruggen2013}, extending the interrogation time in Ramsey experiments \cite{PhysRevX.5.021011} and feedback cooling of a spin ensemble~\cite{Behbood2013}. 


To fully exploit the potential of ultracold clouds in emerging quantum technologies, these atomic samples must be prepared with unprecedented precision. {For instance, precise atom number preparation is crucial to improve the precision of atomic clocks, which is presently limited 
by collisional shifts ~\cite{Ludlow2015}. It is of particular relevance for  techniques that employ interactions to produce non-classical states for improved interferometric sensitivity~\cite{Gross2010,Riedel2010,Lucke2011,Strobel2014}. In general, if  the  number fluctuations  of an atomic ensemble in a single spatial mode can be suppressed, the many-particle state becomes non-classical, yielding a resource  for  atom  interferometry  beyond  the standard  quantum limit~\cite{PhysRevLett.110.163604}. } 
Sub-Poissonian preparation of micro- and mesoscopic atomic samples was recently demonstrated by using single-site addressing in an optical lattice~\cite{Zeiher2015}, 3-body collisions~\cite{Whitlock2010,Itah2010}, non-destructive measurements of nanofiber-based systems~\cite{Beguin2014}, and careful tailoring of the trapping potential for fermionic~\cite{Serwane2011} and bosonic systems~\cite{Chuu2005}. However, despite initial attempts towards the compensation of number fluctuations in ultracold atomic clouds~\cite{Sawyer2012}, the high precision preparation of large atom numbers remains an unsolved challenge.


In this Letter, we stabilize the atom number in ultracold  clouds through the real-time analysis of dispersive images and  feedback, as shown in Fig.~\ref{fig1}(a). 
After initial evaporative cooling of an atomic cloud, a first set of non-destructive Faraday images, ``F1'', determines the number of atoms. We characterize this imaging method and show  it  achieves an atom number uncertainty  below the shot noise level. Based on the analysis of the images, feedback is applied to reduce the atom number to a user-defined target, whereupon a second imaging series,``F2'', probes the remaining number of atoms in the cloud. We show that this technique can stabilize the atom number below the shot noise level.

The evolution of the atom number distribution throughout the sequence can be understood as follows.
The high precision of the Faraday imaging sets the width of the atom number distribution at F1  below the shot noise level.
  To stabilize the atom number, a precise fraction of the atoms is  removed from the cloud, causing the width of the atom number distribution to grow. 
   In general, the loss process between F1 and F2 can be modeled by a master equation, the solution of which yields the probability distribution for the number of trapped atoms as a function of time~\cite{vanKampen}. For single particle loss, however, the atom number distribution is Poissonian, which  may be approximated by a binomial distribution for large $N$ due to the central limit theorem.
This motivates the following simplified model: Starting with  $N_0$ atoms in the trap, where each atom has a probability $p$ of remaining trapped, the number of remaining atoms has a binomial distribution $N \sim B(N_0,p)$, with mean value $\mean{N} = N_0p$ and variance $N_0p(1-p)$.
Thus, the relative uncertainty of the number of atoms remaining in the cloud is given by $\sqrt{(1-p)/\mean{N}}$. From this simple analysis, it is clear that for low levels of applied loss ($p\simeq 1$),   samples with a relative uncertainty well below the shot noise level $1/\sqrt{\mean{N}}$ can be prepared, provided the feedback does not add additional noise.

Having outlined the sequence to stabilize the atom number, we now give a more detailed description of the experiment. Figure~\ref{fig1}(b) shows a schematic of the key components in the experimental setup. Ultracold atomic clouds are produced by forced radio-frequency (RF) evaporation in a Ioffe-Pritchard magnetic trap. The trap has radial and axial trapping frequencies of \mbox{$\omega_\rho=2\pi\times296$~Hz} and \mbox{$\omega_\mathrm{z}=2\pi\times17.1$~Hz,} at a 330~mG bias field. The RF evaporation is stopped at a frequency of 1900~kHz, yielding on average $6.7\times10^6$ \Rb atoms at 18~$\mu$K in the $\ket{F=2,m_F=2}$ state. 

The dispersive imaging employs off-resonant light pulses propagating along the $z$ direction, with an initial polarization along $y$. The Faraday effect leads to a rotation of the linear polarization by an angle $\theta\propto\tilde{n}$, with $\tilde{n}$ the  column density~\cite{Gajdacz2013}. The rotated light is sent through a polarizing beamsplitter (PBS) and imaged on an  Electron Multiplying Charge Coupled Device (EMCCD) camera. This configuration realizes a  ``dark field'' dispersive imaging technique~\cite{Bradley1997,Gajdacz2013}. \del{in which the non-scattered light is removed from the imaging path by the PBS.} The Faraday imaging sequence is realized with light that is blue detuned by $\delta=2\pi\times(1200\pm1)$~MHz from the $F=2\rightarrow F'=3$ transition. Over the spatial extent of the cloud, the intensity distribution is approximately uniform at a value of  0.5~mW/cm$^2$. The imaging light is monitored on a photodiode (PD) on the reflecting port of the PBS  and based on this signal the imaging power is stabilized.   F1 and F2 contain 50 and 100 rectangular pulses, respectively,  with a cycle period of 7~ms. For each   pulse, an image is acquired on the  camera.

These images are  evaluated in real-time on a field programmable gate array (FPGA), which calculates the fraction of atoms to be removed. To apply  feedback, the FPGA controls a synthesizer to generate RF pulses that induce the desired loss. A 10~s delay between F1 and F2 allows time for the loss pulses to be applied and for the cloud to thermalize. In the absence of applied loss, the cloud contains on average $4.3\times10^6$ atoms at 10~$\mu$K after F2. The temperature and atom number at F2 is the combined result of  free evaporation and single particle loss due to finite background pressure. Following F2, the  trap is extinguished and the cloud is probed by resonant absorption imaging after 10~ms time-of-flight.

\begin{figure}[t]
  \includegraphics[width=8.6cm]{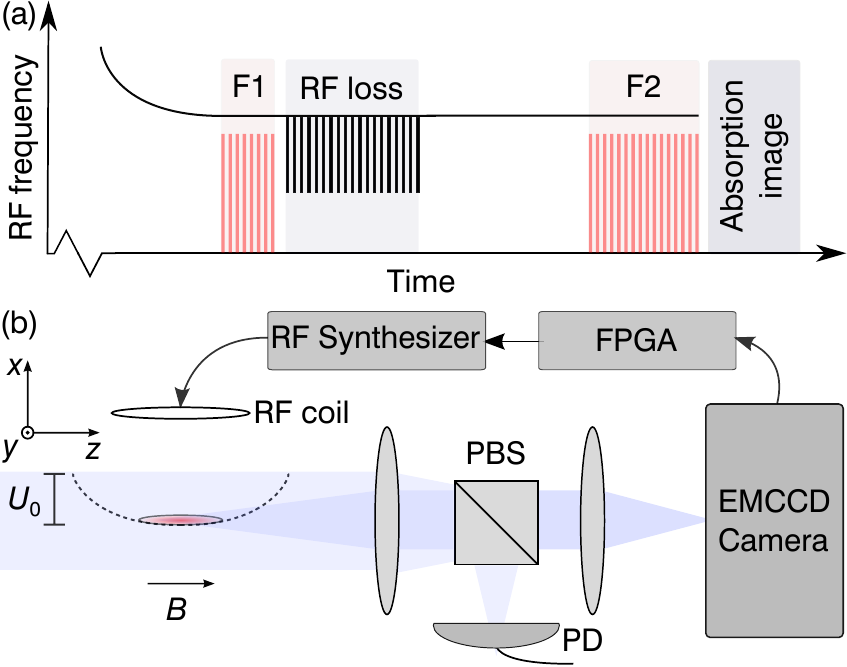}
	\caption{(Color online)  (a) Experimental sequence. 
	An evaporatively cooled cloud is probed by two series of Faraday images, where RF loss pulses between  F1 and F2 remove a controllable fraction of the atom cloud.
	(b) Experiment schematic. Faraday imaging probes a cold atom cloud held in a magnetic trap. The images are processed  in real-time and the outcome can be used to determine the fraction of atoms removed by RF loss, thereby producing a number stabilized cloud.}
	\label{fig1}
\end{figure}

In the following, we evaluate the precision attained by Faraday imaging and then characterize the applied loss mechanism. Based on these results, we  characterize the correlation between the measurements at F1 and F2, since this sets the limit for the precision that can be obtained by feedback. Finally, we show that atomic clouds below the shot noise level can be prepared by our feedback technique.

In the dark field Faraday imaging method, the light intensity on the camera scales with the ``signal'', defined as \mbox{$S \equiv \sin^2 \theta$.} We calculate this signal experimentally by \mbox{$S = \left(I(\theta)/I_\mathrm{ref} - 1 \right) \CS$,} where $I_\mathrm{ref}$ is the intensity of the non-rotated light that leaks through the PBS due to its finite extinction ratio (``cube suppression'') $\CS\sim 10^{-3}$~\cite{Gajdacz2013}. 
The reference intensity $I_\mathrm{ref}$ is obtained from a region outside the atomic signal and the baseline level of the camera is removed from $I(\theta)$ and $I_\mathrm{ref}$ by analyzing a masked region of the camera chip ~\cite{SM}. This procedure makes $S$ independent of the EMCCD gain, which is prone to drift. The rotation angle, and hence the atomic density, can  be obtained from $\theta=\arcsin(\sqrt{S})$, but, in practice, this is complicated by detection noise where $\theta$ is small, leading to negative values of $S$.


To avoid such technical issues on the FPGA, we calculate the signal sum $\Sigma_S$ by summing $S$ in a region-of-interest that encompasses the cloud. {In the limit of small Faraday rotation angles, $S\approx\theta^2$, yielding $\Sigma_S \propto N^2/T$ for the thermal clouds in this work. 
We have characterized the  scaling of $\Sigma_S$ with $N$ and $T$ using results from absorption imaging. We find the observed functional dependence is well described by an empirical model motivated by the small angle dependence of $\Sigma_S$ and that, to a good approximation, the fluctuations in temperature can be neglected ~\cite{SM}. Due to the quadratic dependence of $\sigsum$ on $N$ in the small angle limit, the relative fluctuations in the signal sum, $\Delta\Sigma_S/\sigsum \approx 2\Delta N/N$, are approximately twice as large as those in $N$, making it a sensitive atom number probe.  This approach allows us to exploit the high precision of  Faraday imaging in combination with the accuracy of  absorption imaging to determine the atom number.}

The precision of this Faraday imaging technique can be obtained from an analysis of the fluctuations in $\sigsum$. Figure~\ref{fig2}(a) shows $\sigsum$ at F1 as a function of image number using an imaging pulse duration of $t=0.66$~ms. The signal sum decreases over the 50 imaging pulses as a result of atom loss, primarily due to spontaneous scattering into untrapped electronic states.
 The fluctuations of the signal sum about this mean decay correspond to the light shot noise, the stochastic noise arising from atom loss, and potential technical noise. Since the mean atom loss is deterministic, one can remove the decay by normalizing $\sigsum$ with its mean over several experimental runs and shift it to be centered on zero. Figure~\ref{fig2}(b) shows this normalized signal $E\equiv\sigsum/\langle \sigsum \rangle-1$, which we call the ``error''. {We use the mean value of each error trace  as a measure of the atom number in a given experimental run.}

To characterize the imaging noise, we calculate the two-sample deviations $\Delta\Sigma$ and $\Delta E$ of $\Sigma_S$ and $E$, respectively, for each trace. The relative uncertainty of $\Sigma_S$ in a single imaging sequence is then given by the mean of $\Delta\Sigma$ over $M$ imaging pulses, $\sigma_\Sigma=\frac{1}{\sqrt{M-1}}\expec{\frac{\Delta \Sigma}{\sigsum}}$ and equivalently for $E$ it is \mbox{$\sigma_E=\frac{1}{\sqrt{M-1}}\expec{\Delta E}$}. In total, the relative uncertainties are given by the mean value over several experimental runs denoted by $\expec{\sigma_\Sigma}$ and $\expec{\sigma_E}$.

\begin{figure}
	\centering
		\includegraphics[width=8.6cm]{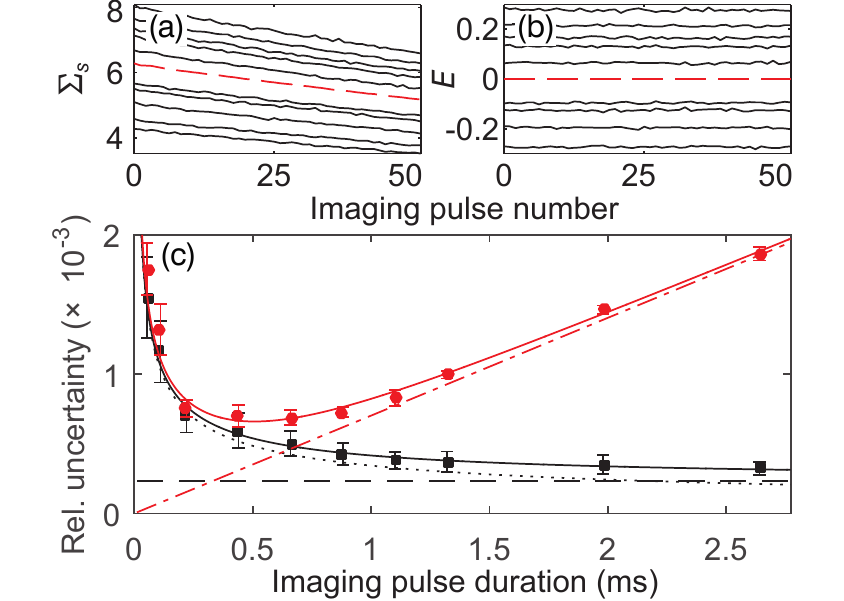}
	\caption{(Color online) (a) Signal sum $\Sigma_S$ v. image number for F1.  (Black lines) Traces for 10 representative runs with imaging pulse duration  $t=0.66$~ ms, (red dashed line) mean trace $\mean{\Sigma_S}$ for the entire data set. (b) Error $E$ 
	 for the same runs as (a). 
	(c) Relative uncertainty v. $t$. (Red circles) $\expec{\sigma_\Sigma}$, (red line) fit of imaging model $\sigma_\textrm{Mod}$, (red dash-dot line) mean atom loss contribution to $\sigma_\textrm{Mod}$ fit $\propto t$. (Black squares) $\expec{\sigma_E}$, (black line) fit of $\sigma_\textrm{Mod}$, (black dotted line) light shot noise contribution to $\sigma_\textrm{Mod}$ fit $\propto 1/\sqrt{t}$ , (black dashed line) constant technical noise contribution to $\sigma_\textrm{Mod}$ fit. Error bars denote the standard error of the mean over several experimental runs.
		}
	\label{fig2}
\end{figure}

Figure~\ref{fig2}(c) shows the relative uncertainties of the signal sum, $\expec{\sigma_\Sigma}$, and the error, $\expec{\sigma_E}$, as a function of imaging pulse duration $t$. Initially, $\expec{\sigma_\Sigma}$ decreases as the imaging pulse duration is increased, but at 0.5~ms it reaches a minimum and then increases approximately linearly with~$t$. The relative uncertainty $\expec{\sigma_E}$ shows the same initial behavior, but does not  increase for the range of $t$ we consider. {To understand this scaling, we use the following noise model~\cite{[{See~}][{ and its supplemental material.}] Hume2013}: \mbox{$\sigma_\textrm{Mod}=[At^{-1}+Bt+Ct^2+D]^\frac{1}{2}$.} 
 The first term is the variance due to   light shot noise; the second  and third terms describe stochastic  and mean atom loss, respectively;  finally, the constant  term $D$ represents technical noise sources in the EMCCD,  uncertainty in the imaging light detuning, and noise arising from the  evaluation of $S$. 
  The loss parameters are closely linked because they describe two effects of single atom loss: In the limit of low loss,  $C= 2B^2$, where $Bt=N_0 (1-p)/2$, and since $N_0\sim10^6$ and $p\simeq1$, the noise term $Bt$ describing stochastic atom loss is negligible compared to $Ct^2$.
Fits to $\expec{\sigma_\Sigma}$ and $\expec{\sigma_E}$ 
are shown in Fig.~\ref{fig2}(c). The fit to $\expec{\sigma_\Sigma}$ is dominated by the light shot noise and mean atom loss terms; this is characteristic of the two-sample deviation in an imaging method that induces significant atom loss~\cite{Hilliard2015}. The fitted values of $A$ and $C$ are consistent with estimates obtained from direct evaluation of the images.
In contrast, $\expec{\sigma_E}$ is well fitted by only the light shot noise and technical noise, since the mean atom loss contribution has been removed by normalization.} Indeed, for $t\lesssim1$~ms, $\expec{\sigma_E}$ is approximately equal to the light noise, a fact we will employ in the following. At the optimal pulse duration of $t=0.66$~ms, the relative uncertainty of $E$ is $5\times10^{-4}$, which yields a relative uncertainty in the detected atom number of  $2.5\times10^{-4}$. The imaging shows a similar performance at F2, for which the optimal imaging pulse duration is 0.55~ms. Thus, the Faraday imaging technique allows us to determine the atom number for clouds containing $\sim 5\times10^6$ atoms approximately a factor of two below the atom shot noise level.

To perform  feedback, we require a mechanism to remove atoms from the cloud. It is important that this mechanism provide sufficiently fine resolution and does not drastically alter other parameters of the system such as the cloud's temperature. A convenient loss mechanism is realized by applying a variable number of fast RF pulses: These pulses transiently lower the trapping potential~\cite{Campo2008}, whereby atoms are lost due to spilling. In general, we employ an RF frequency corresponding to 95\% of the trap depth $U_0$, with a pulse duration of $8.4~\mu$s repeated every $50.4~\mu$s. 
Since the elastic collision rate throughout the experiment after F1 is $\sim100$~Hz, this pulse duration is short compared to the mean time between collisions. These parameters are chosen to achieve a very small fractional loss of $\sim10^{-5}$ per pulse, thus providing fine digital resolution. For example, to remove 10\% of the atoms, we apply  $\sim10^4$ loss pulses.

Based on these results, we characterize the level of correlation between the measurements at F1 and F2. The fluctuations of this correlation set the limit for the precision that can be achieved with feedback, since the feedback strength for a  desired result at F2 is calculated from the signal obtained at F1. The correlation is measured for several fixed applied loss settings. Figure~\ref{fig3}(inset) shows the outcome of such a measurement in terms of the mean measured error at F1 and F2, where the error varies by $\pm40\%$ due to the natural fluctuations of the experiment; this corresponds to a $\pm20\%$ span in atom number. To evaluate how well E1 and E2 are correlated, we  fit a quadratic function to the data and subtract this fit from the data. The relative fluctuations between E1 and E2 are then determined by taking the two-sample deviation over successive runs of the experiment to remove slow drifts in the apparatus such as changes in the trap bottom.

Figure~\ref{fig3} shows the relative uncertainty in the detected total signal at F2 for a number of fixed loss settings, corresponding to several mean numbers of atoms remaining in the trap. This correlation data allows for an analysis of the inherent noise sources. The data is well described by contributions from the imaging noise and from the stochastic noise due to the atom loss between the two imaging series. The  imaging noise  is given by the light shot noise and technical noise contributions in F1 and F2 (corresponding to $\expec{\sigma_E}$ for each imaging series) added in quadrature. It has been fitted by a function $\propto 1/N$, which we expect from error propagation ~\cite{SM}.  For the stochastic atom loss contribution, the relative uncertainty in the number of atoms remaining in the cloud $\sqrt{(1-p)/\mean{N}}$ is shown, which has been transformed into the signal sum using the model linking $\sigsum$ to $N$ and $T$. The total noise, given by the quadrature sum of these contributions shows very good agreement with the experimental data. This crucial result shows that there are no unknown technical noise sources that influence the number of atoms between F1 and F2, which is a  prerequisite to perform feedback below the shot noise level.

\begin{figure}[t]
    \includegraphics[width=8.6cm]{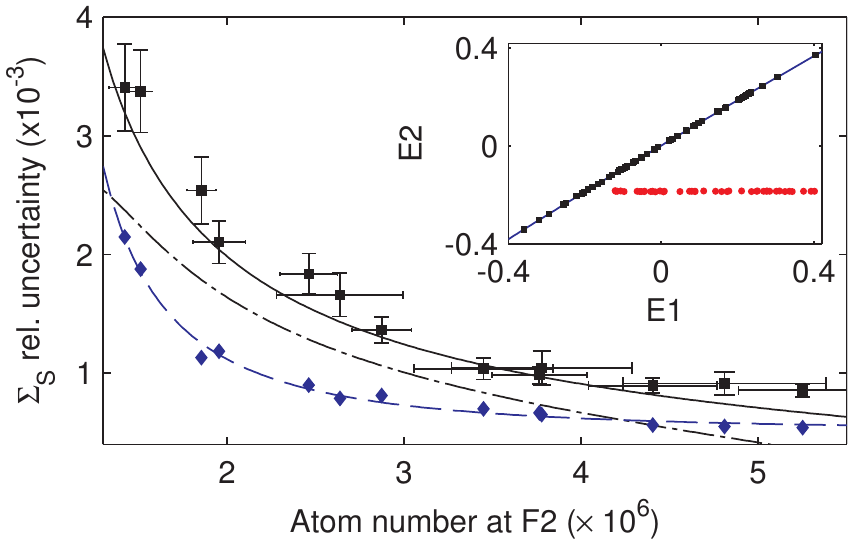}
		\caption{(Color online)   (Black squares) Signal sum relative uncertainty v. atom number at F2, (black line) total noise model, (blue diamonds) imaging noise contribution, (blue dashed line)  fit $\propto1/N$ to imaging noise data, (black dash-dot line) stochastic atom loss noise contribution $\sqrt{(1-p)/\mean{N}}$. Error bars  obtained by bootstrapping the data set.
		Inset: Correlation of error at F1 and F2. (Black squares) Correlation data with no applied loss, (blue line) quadratic fit to correlation data, (red  circles) feedback stabilized points.}
	\label{fig3}
\end{figure}

The achievements outlined above allow us to turn to the active stabilization of the atom number by feedback. Based on the average error in F1, the fraction of atoms removed from the cloud is controlled by varying the number of applied loss pulses. To generate a reference signal $\langle\sigsum^\mathrm{ref}\rangle$ for this feedback, we typically cycle the experiment with no applied loss for $\sim 50$ runs. In subsequent experimental runs with feedback, E1$^\prime \equiv \sigsum/\langle \sigsum^\mathrm{ref} \rangle -1$ is determined at F1, and the number of RF loss pulses $N_\textrm{Loss}$ is calculated using a cubic feedback function   $N_\textrm{Loss} = gE1^\prime(1 + q E1^\prime + c E1^{\prime2}) + d$. This function approximates the atom loss that is exponential in the number of applied loss pulses. The linear $g$, quadratic $q$ and cubic $c$ gain parameters as well as the offset $d$ are determined by evaluating the outcome over several experimental runs for a trial set of feedback parameters and iterating ~\cite{SM}.  Figure~\ref{fig3}(inset) shows a data set where the feedback parameters have been optimized to achieve a stabilized value of E2 for all initial errors E1$^\prime$ that are larger than the target value. In this case, the stabilized atom number is $\sim 90\%$ of the mean atom number of the free running experiment.

Finally, the uncertainty in the stabilized sample is characterized to verify that the feedback mechanism does not add additional noise and that stabilization below the shot noise level can be achieved. 
We take the two sample deviation of E2 over successive runs of the experiment for several target atom numbers. The relative uncertainty is shown as red circles in Fig.~\ref{fig4}. For clouds prepared at $N\gtrsim2.5\times10^6$, the feedback achieves a level of stabilization that is limited by the fundamental noise imposed by the single particle loss mechanism,  showing that the feedback does not induce additional noise. These clouds are stabilized at or below the atom shot noise level $1/\sqrt{N}$. 

\begin{figure}[t]
    \includegraphics[width=8.6cm]{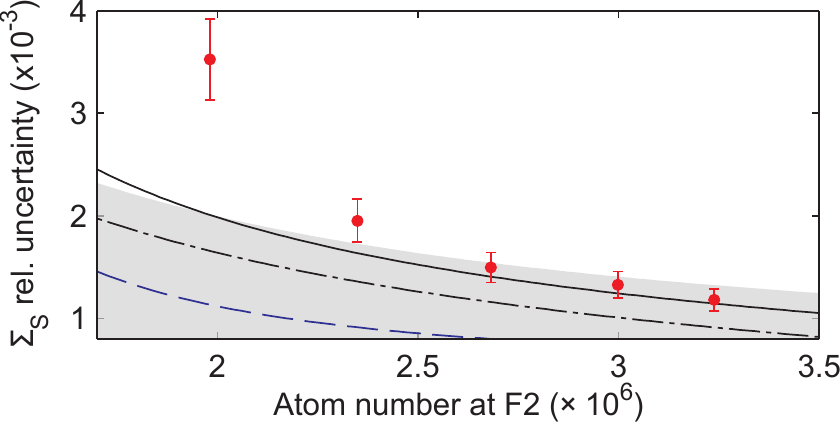}
	\caption{(Color online)  (Red circles) Signal sum relative uncertainty v. atom number at F2 for feedback stabilized clouds.  Other lines same as Fig.~\ref{fig3}. (Shaded region) Interval where the relative uncertainty in the signal sum lies below the shot noise level, i.e., $1/\sqrt{N}$ expressed as signal sum ~\cite{SM}.}
	\label{fig4}
\end{figure}

For samples stabilized to lower atom numbers, the observed noise exceeds the correlation data and the noise model. For $N<2.5\times10^6$, we remove more than 60\% of the atoms from the cloud, whereby the modeling of the exponential loss by the cubic feedback function becomes less accurate. Additionally, the passive stability of the apparatus, such as drifts in the trap bottom, becomes a significant source of noise in the stabilized atom number for high fractional loss.

In conclusion, we have prepared number stabilized atom clouds through feedback.
An investigation of our non-destructive imaging technique yielded an uncertainty in the measured atom number that was about a factor of two smaller than the atom shot noise level. 
The precision of correlation measurements within an experimental realization was entirely determined by the removed fraction of atoms, demonstrating the absence of  technical noise sources between the imaging series. Finally, feedback based on a non-destructive measurement allowed for  stabilization at or below the level of $1/\sqrt{N}$ for large atom clouds with \mbox{$N\gtrsim2.5\times10^6$.} 

{The potential of our technique can be further exploited  by employing multiple feedback steps and improved atom number determination. A second feedback step requiring only a small removal of atoms would strongly reduce the induced noise, whereby the imaging noise would become the limiting factor. To improve the Faraday imaging, more sophisticated atom number estimators will be used to better exploit the information from multiple images. Additionally, the atom number decay due to imaging itself could realize the final feedback step~\cite{Hilliard2015}, in which case an algorithm such as a Kalman filter would stop the imaging series when the target atom number is detected.
However, even at the present level, our technique can make a considerable contribution to improve the precision of current~\cite{Ludlow2015} or non-classical~\cite{Gross2010,Riedel2010,Lucke2011,Strobel2014} atom interferometers.}

\pagebreak
\widetext
\begin{center}
\textbf{\large Preparation of ultracold atom clouds below the shot noise level: Supplemental Material}
\end{center}
\section{Image acquisition and analysis}\label{sect:image}

The Faraday images are acquired on an Andor iXon DU-888 EMCCD camera. The camera is triggered by the experimental control system and, when the exposure is complete, the image pixels are read out at 10~MHz through the camera output amplifier with 14~bit resolution. The electron multiplying gain of the camera permits the amplification of weak light signals above the camera read-out noise. This allows for a good signal-to-noise ratio at high detuning and low intensity which limits the atom loss induced by the Faraday imaging. As a result, we can take multiple images of the same cloud, allowing us to enhance the precision of the measurement through averaging. Additionally, we can estimate the imaging precision by analyzing the fluctuations across the images in a given series.

To analyze the images and perform feedback, we use a National Instruments PCIe-7852R FPGA programmed in LabVIEW. A digital repeater installed on the cable from the camera to the analysis computer allows us to tap into the data transfer and feed it into the FPGA. 
%

\begin{figure}[b]
	\centering
\includegraphics[width=8.6cm]{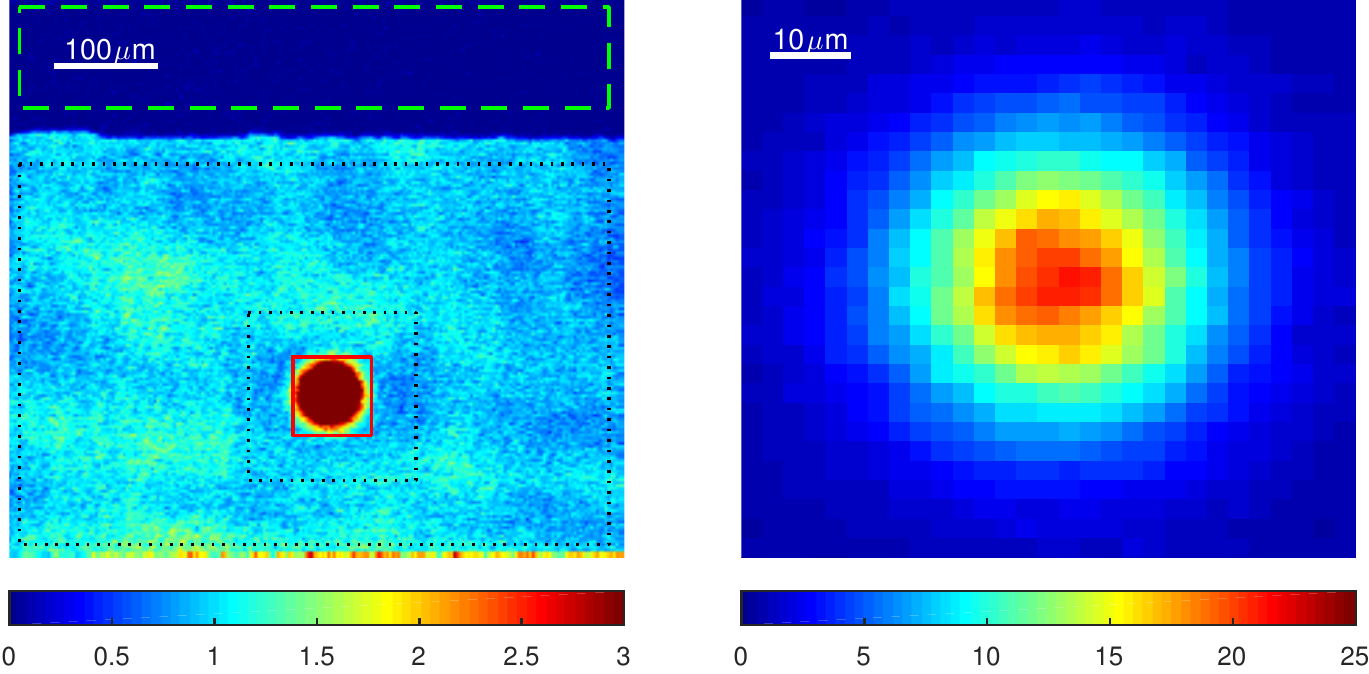}
	\caption{Example of a Faraday image. The left frame shows the full captured $220 \times 200$ pixel frame. The central $29\times29$ pixel region (solid red square) that encompasses the atomic cloud is used to obtain $I(\theta)$. The area between the two dotted black rectangles is used to estimate the reference intensity $I_\textrm{ref}$ from the light that leaks through the PBS. The mean pixel count within the dashed green rectangle corresponds to the offset level and is subtracted from $I(\theta)$ and $I_\textrm{ref}$. The right frame shows the rescaled Faraday image $I(\theta)/I_\textrm{ref} \propto S$.}
	\label{fig:exampleFarImg}
\end{figure}
The signal $S$ is obtained as follows from the three preselected regions-of-interest (ROIs) shown in Fig.~\ref{fig:exampleFarImg}. 
The pixel values recorded in the central ROI correspond to $I(\theta)$. The reference ROI provides the mean intensity $I_\textrm{ref}$ from the light that leaks through the PBS (see Fig.~1(b) main text). The offset ROI corresponds to an area of the chip that is masked by a razor blade in an intermediate image plane. It provides a mean intensity corresponding to background light intensity and the camera offset which is subtracted from $I(\theta)$ and $I_\textrm{ref}$.

Based on these values, $\Sigma_S$ is calculated by summing $S$ in the central ROI. This value is stored in the FPGA within an imaging sequence and used to calculate the desired feedback. The full captured $220 \times 200$ pixel frame is also read out to the camera computer, allowing for post evaluation. With this frame size, the minimal time interval between images is $7\mathrm{ms}$ for continuous imaging, which is limited by the camera readout speed.

\section{Signal noise model}\label{sect:noise}

The correlation of the F1 and F2 measurements is limited by two fundamental sources of noise: the signal sum measurement precision and the stochasticity of the loss between the two measurements.
As we show in Fig.~2 (main text) and the associated discussion, the measurement precision of the signal sum for the chosen imaging parameters evaluated is dominated by the light shot noise.
The stochastic noise due to atom loss leads to a relative uncertainty $\sqrt{(1-p)/\langle N\rangle}$ in the trapped atom number.
{To describe the stochastic noise arising from the loss between F1 and F2, and to include the  contribution to the imaging induced stochastic noise when averaging over F1 and F2,} we approximate the survived fraction by 
$p \approx \langle N_2\rangle/\langle N_1\rangle$, where $\langle N_{1}\rangle$, $\langle N_{2}\rangle$ is the mean atom number in the F1, F2 imaging series, respectively.
The expected stochastic noise is then 
\begin{equation}
\sigma_N 
\approx\sqrt{\frac{1 - \langle N_2\rangle/\langle N_1\rangle}
	{\langle N_2\rangle}}.
\label{eq:expectedPoisNoise}
\end{equation}

We now  derive how the stochastic atom number fluctuations propagate into in the signal sum. The  function $\Sigma_S(N,T)$ linking the atom number and temperature from absorption images to the signal sum  can be obtained by fitting a model to  the fixed applied loss data shown in Fig.~3 (main text).
%
%
\begin{figure}[h!]
	\centering
\includegraphics[width=8.6cm]{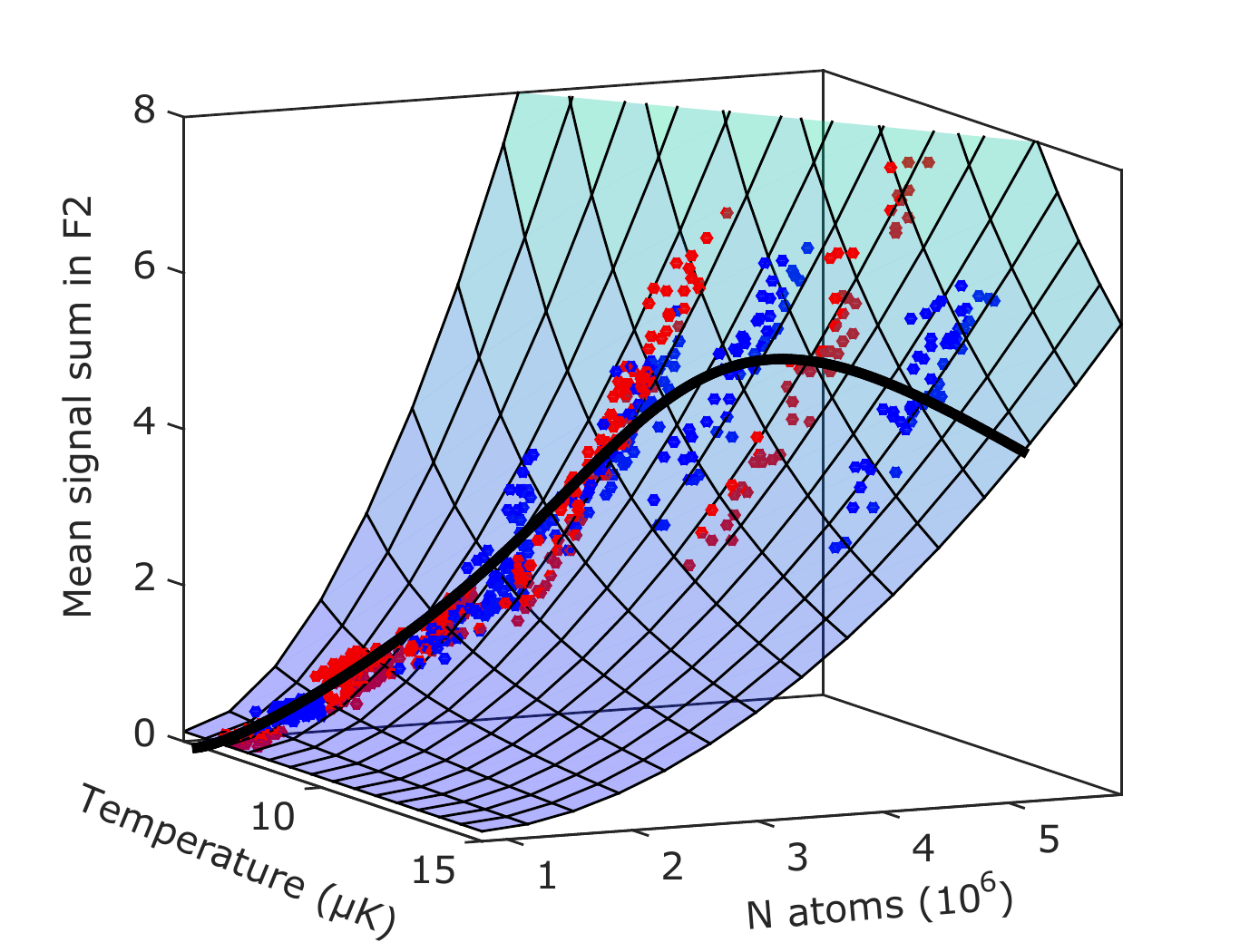}
	\caption{Characterization of the signal sum function 
		$\Sigma_S \equiv \Sigma_S(N,T)$  using  the fixed loss data sets (see Fig.~3 main text). (Alternating red and blue filled circles to distinguish the different data sets) Signal sum v. the atom number $N$ and  cloud temperature $T$  extracted from absorption images. The surface mesh is a fit  of Eq.~(\ref{eq:SNTfitFunc}) to the data. (Black line)  Path on the surface along which the error propagation coefficient in Eq.~\eqref{eq:gammaNanalytical} is evaluated.
		} 
	\label{fig:SNTfunc}
\end{figure} 
The expected functional expression for $\Sigma_S(N,T)$ is motivated by the small angle dependence of the signal $S \approx \theta^2$. 
The rotation angles are Gaussian distributed for a thermal cloud, yielding $\theta\propto N/T$.
Integrating $S$ over the cloud area  gives $\Sigma_S(N,T) \propto N^2/T$. To allow for a more general dependence that is not constrained to small values of $\theta$, we use the following function
\begin{equation}
\Sigma_S(N,T) = a_1 \frac{(N-a_5)^{a_2}}{T^{a_3}} + a_4.
\label{eq:SNTfitFunc}
\end{equation}
The fitted surface is shown Fig.~\ref{fig:SNTfunc}.
The parameter $a_4$ accounts for inhomogeneity in the Faraday beam profile, which gives rise to an offset in the signal sum ROI. The parameter $a_5$ models a lower bound in the measurement of small angles; this arises from the finite extinction ratio of the polarizing beamsplitter in the imaging system (see Fig.~1(b) main text) and it primarily affects the wings of the cloud where the density and hence rotation angles are low  \cite{Gajdacz2013}. The fitted parameters are $a_1 = (18.3\pm1.6)\times 10^{-18}$, $a_2 = 1.82\pm0.03$, $a_3=1.51\pm0.03$, $a_4 = 0.13\pm0.05$ and $a_5 = (8.0\pm0.7)\times10^5$. 

It is evident that the cloud temperature varies in a correlated manner with the atom number at F2  for each fixed applied loss dataset  in Fig.~\ref{fig:SNTfunc}. This is caused by the coupling of $N$ and $T$ in  evaporative cooling \cite{Ketterle1996181}.
We observe, however, that the relative variation in $T$ is approximately a factor of ten smaller than the relative variation in $N$.
Thus, fluctuations in $T$ induced by the stochastic loss of atoms are a factor of ten smaller than the atom number fluctuations. We evaluate the error propagation coefficient for $T$ to be about two times lower than that for $N$, so that the effect of temperature fluctuations on the signal sum is $\sim 20$ times lower. Thus, we neglect  temperature fluctuations in our simple noise model.

By  error propagation, the relative atom number fluctuations propagate into the signal sum error as 
\begin{equation}
\sigma_{\Sigma_S} =  
\frac{\partial \Sigma_S}{\partial N}
\frac{\langle N\rangle}{\langle \Sigma_S \rangle} \sigma_N \equiv \gamma_N \sigma_N,
\label{eq:stochNoiseErrProp}
\end{equation}
where we have defined the error propagation coefficient $\gamma_N$. Using the expression for $\Sigma_S(N,T)$  in Eq.~\eqref{eq:SNTfitFunc}, the error propagation coefficient can be evaluated as
\begin{align}
	\nonumber
\gamma_N(N,T) &= \frac{\partial \Sigma_S}{\partial N} \frac{ N}{\Sigma_S} = a_2 a_1 \frac{(N-a_5)^{a_2-1}}{T^{a_3}} 
\frac{ N}{\Sigma_S} \\
&=  a_2 \left( 1 - \frac{a_4}{\Sigma_S(N,T)} \right)  \left(1 - \frac{a_5}{N}\right )^{-1}.
\label{eq:gammaNanalytical}
\end{align}

To complete the stochastic noise model for Fig.~(3)(a) (main text), we require the error propagation coefficient $\gamma_N$ as a function of $\langle N_2\rangle$. 
Using a cubic polynomial, we fit the mean temperature for each fixed loss data set as a function of the mean atom number to obtain the trajectory in the $(N,T)$ plane along which we evaluate Eq.~(\ref{eq:gammaNanalytical}). 
The resulting trajectory is shown in Fig.~\ref{fig:SNTfunc} with a black solid line, and the corresponding error propagation coefficient $\gamma_N$ is shown in Fig.~\ref{fig:gammaN} as a function of $\langle N_2\rangle$. The variation in $\gamma_N$ is caused by the non-zero coefficients $a_4$ and $a_5$. 
For high atom numbers (at concomitant high temperatures), $\gamma_N$  approaches $\sim 2$ as expected from the simple dependence $\Sigma_S \propto N^2/T$. 
The maximum of $\gamma_N$ occurs at $N = 1.3\times 10^6$. In Fig.~3 (main text), we present data in the range $N \in (1.3, 5.5)\times10^6$, and plot the expected magnitude of  noise due to stochastic atom loss using  Eqs.~(\ref{eq:stochNoiseErrProp}-\ref{eq:gammaNanalytical}). Finally, the error propagation coefficient $\gamma_N$ is also used to generate the shaded area representing the atom shot noise limit $1/\sqrt{\langle N \rangle}$ in Fig.~4 (main text).

\begin{figure}[h]
	\centering
\includegraphics[width=8.4cm]{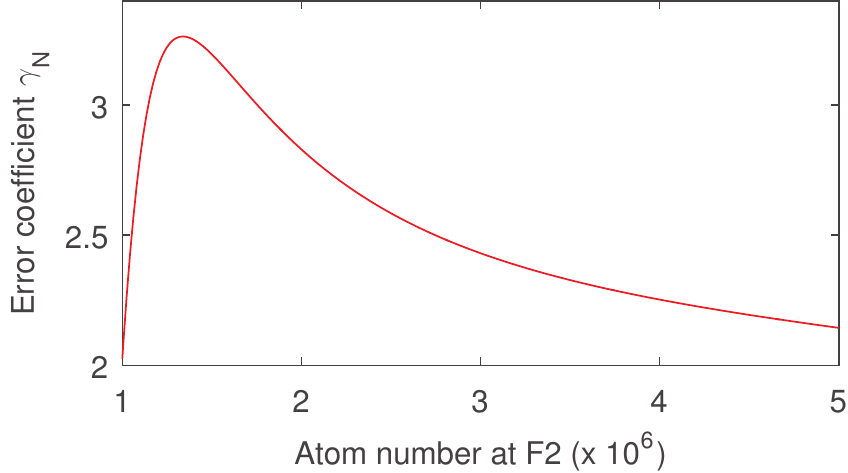}
	\caption{The error propagation coefficient evaluated with Eq.~(\ref{eq:gammaNanalytical}) along the trajectory of mean temperature and atom number.}
	\label{fig:gammaN}
\end{figure}

\section{1/N scaling of the measurement error}\label{sect:measerror}

The average measurement uncertainties of the F1 and F2 series are added in quadrature and plotted in Fig.~3 (main text) (blue diamond symbols). The blue dashed curve is a three parameter fit to the data
\begin{equation}
  \sigma_{\Sigma,\mathrm{meas}} = \sqrt{\left( \frac{b_1}{N-b_2} \right)^2 + b_3^2},
\end{equation}
with $b_1= (1.12 \pm 0.04)\times10^3$, $b_2 = (8.94 \pm 0.09)\times10^{5}$ and $b_3 = (5.7 \pm 0.3) \times 10^{-4}$. The parameter $b_2$ performs a similar function to $a_5$ in Eq.~(\ref{eq:SNTfitFunc}), and, indeed, the two fitted values are  comparable. The parameter $b_3$ models the imaging noise in F1 and any other contributions to the noise that are independent of the atom number. The measurement error grows with decreasing atomic density in the F2 imaging series. The choice of the $\sim 1/N$ fit function is justified in the following.

In reference~\cite{Gajdacz2013}, we showed that signal-to-noise ratio in dispersive imaging techniques induced by the photon shot noise is proportional to the atomic column density. For a fixed temperature of the cloud, the density is proportional to the atom number $N$ and therefore the measurement error scales as $1/N$.

This behavior can also be understood in the following simple model. Assume the number of photons hitting a given camera pixel is
\begin{equation}
  N_{\mathrm{ph}} = N_{\mathrm{ph,0}} S, 
\end{equation}
where $N_{\mathrm{ph,0}}$ is the incident number of photons (on the atoms), and $S\approx\theta^2$. The photon shot noise induced fluctuations of the signal are
\begin{equation}
  \Delta S = \frac{\Delta N_{\mathrm{ph}} }{N_{\mathrm{ph,0}}} = 
  \frac{\sqrt{N_{\mathrm{ph}}} }{N_{\mathrm{ph,0}}} = 
  \sqrt{\frac{S}{N_{\mathrm{ph,0}}}}
  = \frac{\theta}{\sqrt{N_{\mathrm{ph,0}}}}, 
\end{equation}
which induces a relative uncertainty 
\begin{equation}
\sigma_S = \frac{\Delta S}{S} = 
\frac{1}{\theta \sqrt{ N_{\mathrm{ph,0}}}}. 
\end{equation}
As discussed above, $\theta \propto N/T$, thus the measurement error is  inversely proportional to $N$ (for a fixed $T$). 

\section{Choice of the feedback parameters}\label{sect:feedbackparams}

In the feedback experiment, the number of applied loss pulses $N_\mathrm{Loss}$ in each run is determined by the  function
\begin{equation}
  N_{\mathrm{Loss}} = g E1^\prime \left[1 + q E1^\prime + c {E1^\prime}^2  \right] + d,
  \label{eq:feedbackFunction}
\end{equation}
where $g$, $q$ and $c$ are the linear, quadratic and cubic gains, respectively, 
and $d$ is the default number of loss pulses. 
For the application of loss, the value of $E1^\prime$ is evaluated with respect to a mean signal sum from a reference data set (see the main text). 

The feedback parameters are chosen so as to produce a stable value of signal sum in F2, that is set $E2 = \mathrm{const.}$
This is done in an iterative manner.  
First the feedback parameters are guessed, and a trial data set is acquired. In this experiment, the naturally fluctuating $E1^\prime$ samples a range of applied loss pulses $ N_{\mathrm{Loss}}$.  We then make a third order polynomial fit to 
\begin{equation} 
f_{\Sigma_\mathrm{S}} (N_\mathrm{Loss})\equiv \frac{\Sigma_{\mathrm{S,F2}}}{\Sigma_{\mathrm{S,F1}}},
\end{equation}
as a function of the number of applied loss pulses. 
$\Sigma_\mathrm{S,F1}$ is the mean signal sum in F1 and   $\Sigma_\mathrm{S,F2}$ is the corresponding quantity for F2 in a given feedback run. 
For each of the trial runs,
\begin{equation} 
f_{\Sigma_\mathrm{S}}^\mathrm{ideal} \equiv \frac{\Sigma_{\mathrm{S,F2}}^\mathrm{target}}{\Sigma_{\mathrm{S,F1}}},
\end{equation}
 would yield a fixed signal sum $\Sigma_{\mathrm{S,F2}}^\mathrm{target}$.
We can then invert $f_{\Sigma_\mathrm{S}} (N_\mathrm{Loss})$ to obtain the ideal applied loss \mbox{$N_\mathrm{Loss}^\mathrm{ideal} \equiv N_\mathrm{Loss}(
f_{\Sigma_\mathrm{S}}^\mathrm{ideal})$.} The ideal loss can then be fitted as a function of $E1^\prime$ using  Eq.~(\ref{eq:feedbackFunction}), which yields improved values for the feedback parameters. 
Typically, only one to two iterations are required.
Figure~\ref{fig:FeedbackOptimisation} shows  examples of a reference, trial and optimized feedback run.
\vspace{5 mm}
\begin{figure}[h!]
	\centering
\includegraphics[width=8.5cm]{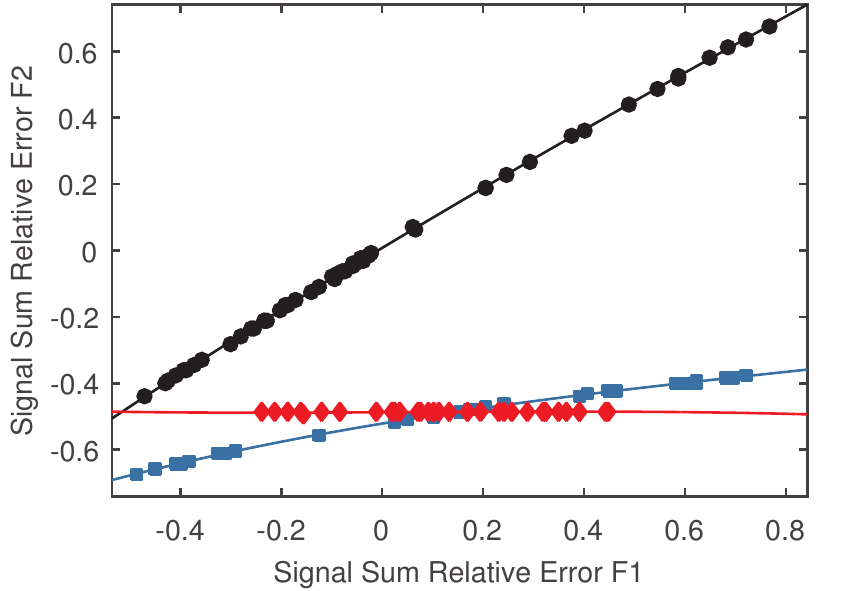}
		\caption{Correlation of signal sum with no feedback (black circles), trial feedback (blue squares) and optimized feedback (red diamonds).} 
	\label{fig:FeedbackOptimisation}
\end{figure}

\section{Statistical uncertainty}\label{sect:stats}

The errorbars on the data points in Fig.~3 (main text) are obtained by bootstrapping~\cite{Efron1979}. This is  independently performed  for each data set (containing on average $\approx50$ runs). After subtraction of the quadratic fit from the correlation of E2 and E1 (see e.g., the inset in Fig.~3 (main text)), the residual deviations from the fit are arranged in the order of their acquisition, and a set of successive two sample differences $\mathcal{D}$ is constructed to eliminate slow drifts in the apparatus contributing to the measured noise. 

For each set $\mathcal{D}$, containing $N_\mathcal{D}$ elements, we randomly draw sets of $N_\mathcal{D}$ samples with replacement and evaluate the standard deviation for each set. The data sampling is repeated 1000 times, which yields the uncertainty of the standard deviation. This result is divided with $\sqrt{2}$ to represent noise in a single experimental realization, which is plotted as vertical errorbars in Fig.~3(main text). 

\section{Acknowledgments}
\begin{acknowledgments}
We acknowledge support from the Danish National Research Foundation, the Danish Council for Independent Research, the European Research Council, and the Lundbeck Foundation. C.K. acknowledges support from  the Centre for Quantum Engineering and Space-Time Research (QUEST) and  the Deutsche Forschungsgemeinschaft (Research Training Group 1729 and CRC 1227).
\end{acknowledgments}

%

\end{document}